# SEMANTIC GROUNDING STRATEGIES FOR TAG-BASED RECOMMENDER SYSTEMS

Frederico Durao[1] and Peter Dolog[1]

[1]Department of Computer Science, Aalborg University, Aalborg, Denmark
`fred@cs.aau.dk`
`dolog@cs.aau.dk`

## ABSTRACT

*Recommender systems usually operate on similarities between recommended items or users. Tag based recommender systems utilize similarities on tags. The tags are however mostly free user entered phrases. Therefore, similarities computed without their semantic groundings might lead to less relevant recommendations. In this paper, we study a semantic grounding used for tag similarity calculus. We show a comprehensive analysis of semantic grounding given by 20 ontologies from different domains. The study besides other things reveals that currently available OWL ontologies are very narrow and the percentage of the similarity expansions is rather small. WordNet scores slightly better as it is broader but not much as it does not support several semantic relationships. Furthermore, the study reveals that even with such number of expansions, the recommendations change considerably.*

## 1. INTRODUCTION

Collaborative tagging has emerged as a useful means to organize and share resources on the Web. In fact, collaborative tagging produces user generated evidence on the interests. Tag based recommender systems utilize this piece of information to find similar resources and generate personalized recommendations. However tags can represent different aspects of the resources they describe and it is not certain whether these tags are sufficient to effectively determine similarities between different resources. Lexical comparison is sometimes not sufficient since tags are personal and variations on its writing may occur. Semantic grounding on tags then opens an optimistic perspective towards more concrete similarity agreements as already pointed in [2].

The identification of semantic relationship between tags establishes solid relations between resources. WordNet dictionary and domain ontologies are utilized in this study to expand the tag meaning in order to establish semantic grounding and the semantic relationship between tags. Nevertheless, the meaning of a tag is highly dependent on the purpose of the author when tagging as known from several studies [4, 12, 8]. Self reference tags, for instance, have particular meaning which hardly may be carried out by generic dictionaries or ontologies. This issue however becomes easier when context they belong to is properly identified. The contextual information can be regarded as the entire information where the tag is assigned to. In this sense, contextual information such as author preferences and sibling tags (i.e. the neighbour tags used to assign the same resource) are key information when determining the meaning of a tag.

In this study, the semantic groundings are explored within three different contexts so that each particular case may result in recommendations which are completely different. Furthermore, we compare the semantic grounding generated using WordNet against ontologies. Data for our study come from three popular social tag-based web systems such as Del.icio.us, Flickr and Digg.

The paper is organized as follows. In Section 2, we describe a motivation scenario for applying semantic grounding. In Section 3, we discuss related work. Section 4 presents our approach to semantic grounding. First, tag meaning expansion is explained in subsection 4.1. Then, subsection 4.2 details the step towards the semantic grounding for tag similarity based on ontology relations. Subsection 4.3 discusses how the user preferences are inferred from tags. Subsection 4.4 illustrates and discusses how recommendations are calculated. Section 5 presents the evaluation analysis of the semantic expansion and grounding for a recommender system with data from social tag-based web systems. Section 6 concludes the study and points out the future works.

## 2. MOTIVATING SCENARIO

Tags from social tag-based system have been recently employed to identify similar resources on the web. In principle, resources which share same tags have high probability of being about the same content and consequently similar. The traditional mechanisms for similarity calculus usually consider the syntax of tags forgetting their meaning. Along this article, the term *resource* will be used as a generic term to refer to a document, video, image, text, file or any sort of asset which can be tagged. Figure 1 shows how basic syntax similarity comparison fails leading to incorrect recommendations. We will use this scenario as an example to illustrate our approach later in this paper.

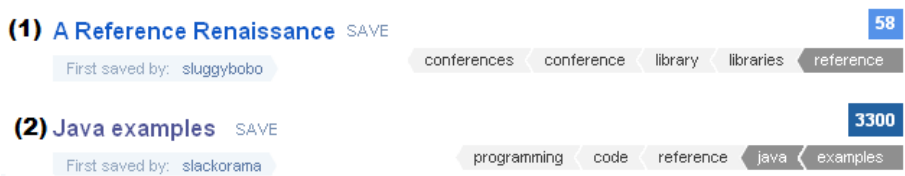

Figure 1.  Similar tag with different meaning

Figure 1 shows two resources from Del.ici.ous which shares the tag "reference". Although this tag is syntactically similar, its meaning varies according to the context it is regarded in. Analyzing the context given by the sibling tags, for instance, we see the resource number 1 is about *conference* whereas the resource number 2 is about *java examples*. When this semantic anomaly is not detected, recommendations which are not relevant to the context or user may be undertaken. Applying semantic grounding on tag similarity measurement represent a forward step against ambiguity problems:

- **Resources have tags syntactically different but similar semantically** - This is a famous case of synonyms and to overcome it some semantic assistance is needed either by use of domain ontologies or looking up for synonyms in dictionary. For instance, tags "plane" and "airplane" looks different but share the same meaning. The obstacle is that generic dictionary sometimes fails to provide the correct meaning of specific terms in a given *context*.

- **Resources share same tags with different meanings** - This is the well known case of polysemy. For instance, the tag "windows" can be about the operating system or the house artefact. Equally to the synonyms, additional semantic support is needed to differentiate their meanings and avoid inconsistent recommendations.

International Journal of Web & Semantic Technology (IJWesT) Vol.2, No.4, October 2011

Although the semantic grounding is a potential method for disambiguation, the quality of the source for semantic grounding in which the meanings are extracted from becomes a key variable to the efficiency of this mechanism.

## 3. RELATED WORK

Tags have been recently studied in the context of recommender systems due to various reasons. [7] argues for a solution where tagging from social bookmarking provides a context for recommender systems in terms of context clues from tags as well as connectivity among users to improve the collaborative recommender system. [10] extends a content based recommender system by deriving current and general personal interests of users from different tags according to different time intervals. [5] develops a page rank based algorithm for recommendations of resources based on preference vectors in folksonomy systems. [16] investigates whether folksonomies might be a valuable source of information about user interests. The main contribution is a strategy that enables a content-based recommender to infer user interests by applying machine learning techniques both on the "official" item descriptions provided by a publisher, and on tags which users adopt to freely annotate relevant items. Static content and tags are preventively analyzed by advanced linguistic techniques in order to capture the semantics of the user interests often hidden behind keywords. [3] shows the benefits of using tag based profiles for personalized recommendations of music on Last.fm. Similar cognition over the product items as subject of recommendations is considered as another factor in addition to the similar tags when personalizing recommendations given by a tag based collaborative recommender system in [13] Similarly as our work, [14] presents the Super Word Set Similarity measure that makes use of WordNet to discover new MCs in ontology matching. The proposed ontology matching method based on the Super Word Set Similarity takes the matching results of the phase of similarity measure between concepts and the phase of similarity measure between properties in order to find new matched concepts. Although the goal of this work is not ontology matching, the WordNet dictionary is utilized for grounding semantic relations. Unlike our method, [15] presents a second order co-occurrence and a related distance measure for tag similarities that is robust against the variation in tags. From this distance measure it is straightforward to derive methods to analyze user interest and compute recommendations.

With the exception of the recent work on Folkrank and semantic grounding of tag relatedness [2], there are not many studies on effect of semantics in tag based systems. As well known from the literature, the tags are just free form keywords used and invented by users to organize their resources. The purpose of tags varies as well as tagging itself may be influenced by different factors. For example, [8] studies a model for tagging evolution based on community influence and personal tendency. It shows how 4 different options to display tags affect user's tagging behaviour. Our study goes beyond this, looking at how existing ontologies change recommendations. [1] studies how the tags are used for search purposes. It confirms that the tags can represent different purpose such as topic, self reference, and so on and that the distribution of usage between the purposes varies across the domains. It compares the purposes with other literature (such as [4, 12, 8] ) where these are called differently.

[9] and [6] coined the term *emergent semantics* as the semantics which emerge in communities as social agreement on tag's meaning based on its more frequent usage instead of the contract given by ontologies from ontology engineering point of view. However, the approaches based on emergent semantics are characterized by the power law which gives a long tail of the tags of which semantics have not emerged yet. Therefore, [2] looks at grounding of the tag relatedness with a help of WordNet.

In this paper we look at, how grounding of relatedness between tags to ontologies including WordNet can have an effect on the recommendations. We wanted to see, whether the ontologies, which are mostly quite specific can be of help in some recommendations and how the grounding

69

International Journal of Web & Semantic Technology (IJWesT) Vol.2, No.4, October 2011or expansions of the set of similar tags by those suggested from ontologies actually change the resulted recommendations. We have not yet found a similar study to this.

## 4. THE SEMANTIC GROUNDING APPROACH

In this paper, we consider the semantic grounding for tag similarity in two steps:
- Tag Meaning Expansion – classes or properties in domain ontologies that are associated with a class that matches a tag are retrieved;
- Tag Similarity Semantic Grounding – subset of retrieved tags are utilized for computing the semantic similarity.

Based on this similarity resolution, we then calculate personalized recommendations by matchmaking to the user preferences inferred from most frequent user's tags.

### 4.1. Expanding the Tag Meaning

The step before performing semantic grounding is to expand the tag in the possible meanings it can assume. Standard meaning can be found in traditional dictionaries such as the WordNet dictionary or similar applications. Nevertheless, ontologies are reasonably necessary when digging for specific domain vocabularies. Both, WordNet and ontologies, when utilized for semantic grounding during the tag similarity calculus, offer a number of semantic expansions, i.e. the meanings which can be raised from a specific tag. In the WordNet dictionary particularly, a tag has a set of cognitive synonyms (synsets), each expressing a distinct concept. Synsets are interlinked by means of conceptual-semantic and lexical relations [11]. In the ontologies, the tag can be found as class concepts or properties associated with it and its understanding depends on the domain regarded by the ontology. The semantic expansions achieved either by WordNet or domain ontologies result in a network of meaningfully related words and concepts that can be used as inputs for semantic grounding.

In general, the semantic expansion achieved from WordNet denotes synonyms or equivalence, on the other hand, the expansions provided by ontologies depends on the meaning given by the properties associated with the concept. A brief example of this can be expressed by Figure 2 which shows semantic expansions from tag "paper".

| Semantic Expansion | Source | Semantic Relationship | Treasure |
|---|---|---|---|
| workshopPaper | Ontology Class | workshopPaper *isSubClassof* Paper | .../concerence.owl |
| conferencePaper | Ontology Class | conferencePaper *isSubClassof* Paper | .../conference.owl |
| handcraft | Ontology Class | Handcraft madeOf Paper | .../arts.owl |
| wood | Ontology Class | paper derivedFrom wood | .../nature.owl |
| report | Synset | paper isSonymyOf Report | WordNet |
| newspaper | Synset | paper isSonymyOf newsPaper | WordNet |
| scientific paper | Synset | scientific paper isSonymyOf paper | WordNet |

Figure 2: Semantic Expansion

This is a clear example which shows that for one single tag, a number of semantic expansions can be carried out and the meaning to be followed impacts directly in the similarity agreement. Another issue is about the quality of the ontologies so that it directly influences the quality of semantic expansions. Either human expertise or automatic mechanism should establish minimal criteria to classify ontology as with high quality or not. Although no formal quality control was

70

International Journal of Web & Semantic Technology (IJWesT) Vol.2, No.4, October 2011

managed in this study, the ontologies utilized here were taken from public spaces on the Web and well known repositories such as Protege[1] and Swoogle[2].

### 4.2. Ontologies Properties Used for Semantic Expansions

Essentially, synsets of WordNet are synonyms which provide definitions of a given input word. A formal representation of this relationship in WordNet could be given as an ontology property which expresses definition, synonym such as "isSynonymOf", "isDefinedAs" or other properties with the same meaning. Differently from WordNet, ontologies have their own semantic network with well defined properties that do not necessarily show synonyms or definitions. However, particularly for this study, only properties which share the same meaning as WordNet will be taken into account because the goal is to identify similar tags from different resources. Then, following the purpose of WordNet, it is reasonable to work with properties which denote definition, synonyms, equivalence, equality and partnership. This reasoning is needed in order to avoid properties which express contradiction or unrelated relationship during the semantic grounding process. Concerning the meaning of the properties as high priority issue, a number of properties were established to be accepted and rejected during the semantic expansion. Table 1 presents some properties utilized for semantic expansion between tag and concepts.

Table 1. Example of some ontology properties considered for semantic expansion.

| Partnership | Equivalence | Definition |
|---|---|---|
| subClassOf | equivalentClass | isA |
| specify | equivalentProperty | hasTypeOf |
| hasPartOf | SymmetricProperty | hasMeaning |
| intersectionOf | sameAs | typify |
| unionOf | similarTo | meaningOf |
| complementOf | associatedWith | belongsTo |
| generalizes | hasRelatedConcept | type |

Properties which do not share the same meaning as those introduced above are automatically rejected because they have different purpose. In addition to native properties obtained from RDF, SKOS, RDFS and OWL vocabulary, a number of data properties were harvested from distinct ontologies on the Web. The ontology properties represent a fundamental step in the process of semantic grounding and at least a minimal reasoning on top of it has to be carried out. Nevertheless, a deeper study on how ontology properties influence semantic similarity calculus can provide worthwhile findings in the way of establishing further groundings.

### 4.3. The Semantic Grounding

The semantic grounding is the concretization of a semantic relationship between two tags from distinct resources. Figure 3 illustrates how the grounding is established.

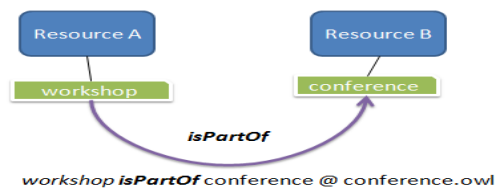

Figure 3: Semantic Grounding

---
[1] http://protege.stanford.edu
[2] http://swoogle.umbc.edu/

*Resource A* has tag *workshop* and *Resource B* has the tag *conference*. The semantic expansion of the tag *workshop* using the ontology conference.owl provides the relationship *workshop isPartOf conference*. At this point, a semantic similarity is found through a relationship which denotes partnership. In this example, only one relationship was found, however, it is common having more.

By looking up the semantic expansions in a formal structured representation of knowledge afford the definition of well-defined metrics of semantic similarity calculus besides reasoning actions. Even though this formalization is fundamental for computational means, the decision on what meaning to follow is the most important step in the way of grounding. The goal therefore is to explore all possible semantic expansions and decide which of them should be used depending on the purpose of the application or recommendation if this is the case. In the following, three strategies for choosing which semantic expansion to follow are discussed.

- **All semantic expansions are considered** – In this case, context is initially ignored and whole set of meanings provided by the semantic expansion are used for comparison. The chances of semantic grounding are high; however, the probability of incorrect grounding increases as well. When the similarity is further applied for recommendation purposes, they have high chance of being about unrelated subject from those indeed expected. Although this methodology is not efficient, it could be viable if combined with some additional mechanism able to identify the rejected recommendations and use this information to gradually refine the next ones.
- **Semantic expansions filtered out by sibling's tags** – The sibling tags can be regarded as hints to precisely determine which expansion to follow. This belief is based on studies in semantic relatedness which claims that sibling tags are about the same content. The absence of sibling tags is the major problem for this methodology because no context is selected and consequently no meaning is used. A possible solution is to switch to the strategy i. (all expansion mechanism) as soon as the lack of sibling's tags is detected.
- **Semantic expansions filtered out by the most frequent tags of the user (MFT)** – Aiming at applying the similarity measurement for personalization means, the semantic expansions can be determined by the top N most frequent tags of the user. Similarly as the sibling tag mechanism, the eminent problem here is the lack of tags able to describe the user preference. Moreover, it is important to state the relatedness between the MFT can be diffuse and the semantic expansion is hardly selected.

To calculate the most frequent tags of a user is not a difficult task, however, to define which of them best represent his/her interest depends on more detailed analysis. The *amount of tags a user has* and the *frequency on which each tag appears* are variables which should be taken into account. Users which contain few tags usually find low frequencies among his/her tags so that it is hard to assure the user's preference. On the other hand, with many tags it is easier to identify the tags which most repeat and then have a clear definition of the preference of the user. Based on this, we have created a formula to define which tags best describe the user preference. In this approach, to faithfully say about interest of a user, we select those tags whose tag frequency is 70% closer to the most frequent tag. In the case on which there are no tags to satisfy this condition, it is assumed the user does not have a clear preference. The objective of this rationale is to guarantee semantic similarity regarding *MFT* as precise as possible.

### 4.4. The Recommendations

In order to illustrate the semantic grounding running under the strategies introduced before, we go back to the Figure 1 and explore how the strategies influence in the generation of the recommendations. Looking up the tag "reference" raises three semantic expansions:

1. reference *isSuperClass* referenceSystem given by context.owl ontology;
2. reference *isTypeOf* JavaDocReference given by java.owl ontology;
3. conference *cotains* reference is given by conference.owl ontology;

Back to the resources, let's take the tag "conference" from the upper resource to compare against "reference" from the inferior resource.

  a) **All semantic expansions are considered** – Following this strategy, the whole set of expansions will be considered and no validation is performed. According to *item 3* it is possible to realize a semantic relation between "reference" and "conference" given by ontology conference.owl. The semantic grounding is established and considering there is no context validation, the semantic similarity is raised incorrectly. If recommendations are motivated by this finding, likely the receiver will not be satisfied and the system has to be provided of intelligent mechanism to maintenance this bad reasoning.
  b) **Semantic expansions filtered out by sibling's tags** – When the context is established by the sibling tags, the semantic grounding is undertaken in two steps. First the semantic relationship between the concepts is found and then the context is assured by the singling tags. The first step is already complete as stated in previously; the second step is then searching for concepts expressed by the sibling tags in the ontology which gave the basis for the semantic grounding. The sibling tags of conference are "library", "libraries" and "conferences". Although there is lexical variations, library and conference are considered in the ontology conference.owl and finally the semantic grounding is established. The semantic similarity is assured more concretely than in the previous strategy and consequently the chances of irrelevant recommendations tend to decrease considerably.
  c) **Semantic expansions filtered out by the most frequent tags of the user (MFT) –** Similarly to the sibling methodology, the semantic grounding is done in two steps: the first is to establish the semantic grounding and the second is to validate the context based on the *MFT* of the user. In the current example, the first step is done but the second depends on whether the user has his preference explicated by tags. If we go to del.ici.ous, it is possible to see that "programming", "java", "howto" and "software" are the top 4 most MFT of the owner of the second resource. After having the MFT information, we realize that the semantic grounding has not been completely fulfilled because the *MFT* has not found as concepts in the context.owl.

It is important to state that none of the strategies can be considered the best. The All Expansion strategy is less restrictive than the others. However, if supported by intelligent mechanisms to detect rejected recommendations, it can work as efficient as the other mechanisms. The *Sibling* and *MFT* strategies are stricter and require further validation of the context. Another possible scenario can be composed by the *Sibling* and *MFT* strategies running together however it is time-consuming and performance issues must be taken into account.

## 5. EXPERIMENTAL EVALUATION

In order to evaluate how semantic grounding strategies impacts the generation of tag-based recommendations, we experimented with three large scale datasets from well known tag-based systems on the Internet: Del.ici.ous[3], Digg[4] and Flickr[5]. In addition, to provide the minimal input for computing semantic similarity, the triple <tag, resource, author>, they were chosen due to a variety of tags motivated by their different natures.

---

[3] http://delicious.com
[4] http://digg.com
[5] http://www.flickr.com

Del.ici.ous is a social bookmarking system in which the assigned tags tend to be more personalized. Flickr is mostly focused on photos and picture sharing whereas Digg is a collaborative sharing website for general content purpose. The three tag-based systems are worth source of tags from the most different domains and give us credibility to be working with real life insights. The overall dataset comprises 6,085 tags and 2491 bookmarks.

## 5.1. The Ontologies Utilized

In order to introduce the ontologies considered in this study, a brief description of the concepts and domains addressed will be presented.

- **biopax-level3.owl** – BioPAX Ontology aims at developing a common exchange format for biological pathway data. It covers metabolic pathways, molecular interactions, signaling pathways (including molecular states and generics), gene regulation and genetic interactions. BioPAX Level 3 is currently under development and review by pathway databases.
- **cancer.owl** – Cancer ontology addresses prognosis, treatment, symptoms and risks of the deasea. In addition, this ontology categorizes a number of variations of cancer according to gender.
- **context.owl** – It describes the entire ambient intelligence for the networked home environment.
- **photography.owl** – An ontology of photography and cameras. Focus on the effect of photographercontrolled factors (such as aperture, shutter speed, equipment used etc) on the properties of the image produced (motion blur, depth of field etc).
- **user.owl** – User ontology addresses user profiling methodology to enhance the effectiveness and usability of services and interfaces in order to tailor information presentation to user and context.
- **country.owl** – Country ontology is about territory issues concerning the delimitations of frontiers or boundaries. Moreover, it addresses political disputes and historical conflicts resumed in current independencies.
- **learning.owl** – This ontology introduces learning concepts involving many aspects which surround a student in the university environment.
- **koala.owl** – This ontology depicts habitats, life conditions, food and scientific classification of Koalas.
- **trip.owl** – This ontology introduces trip issues such as destination, type of accommodation, activities, hotel bookings, sightseeing's and flight arrangements.
- **conference.owl** – Conference ontology is about paper, submission, deadlines and whole vocabulary commonly found in calls for papers. In addition, it categorizes diverse sort of conferences based on their purpose.
- **wine.owl** – This is the classical wine ontology from Protege Web site which categorize a number of variety of wines based on flavor and regions.

In addition to the ontologies described above, ten other ontologies were utilized in the semantic grounding process as well. They are: *java.owl, usability.owl, profile.owl, terrorism.owl, office.owl, animal.owl, resistence.owl, social.owl, sport.owl and food.owl*. Figure 4 shows the domain distribution covered by the utilized ontologies. Although each ontology has a specific purpose, the set of ontologies when analyzed together cover a comprehensive range of domains.

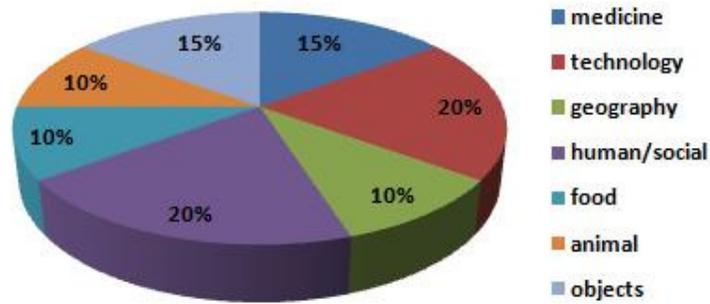

*Figure 4: Domains considered by ontologies utilized*

The following analysis are undertaken with the goal of expressing how semantic grounding impacts in the generation of new recommendations. By comparing the recommendations generated by basic similarity against those processed by the semantic grounding, we have a real notion of how many similar resources have been ignored and should have been considered. In the following subsections we compare the semantic expansions and groundings achieved by 20 ontologies against WordNet. The goal is to measure the variation of the recommendations generated by different treasure. In a sequence, we compare the variation of recommendations processed by basic similarity calculus against those processed by the semantic grounding.

## 5.2. Semantic Expansions and Groundings using WordNet and Ontologies

Table 2. Semantic Expansions and Grounding using WordNet.

| Data | Del.ici.ous | Flickr | Digg |
|---|---|---|---|
| Semantic Expansions | 49% | 63% | 59% |
| All Expansion Strategy | 62% | 48% | 47% |
| Sibling Strategy | 41% | 38% | 53% |
| MFT Strategy | 27% | 44% | 45% |

Table 2 shows the semantic expansions and groundings carried out using WordNet. According to the Table 2, more than 50% of the tags in the three data sets were semantically expanded. The rest of them were not achieved by WordNet either due to syntax variations (e.g *web2.0* and *web2_0*) or intense use of self reference tags (e.g *todo*, *myTopic*). The significant number of expansions is somehow expected because WordNet is a broad dictionary and free of context. Among the three datasets, Del.ici.ous obtained the lowest rate of expansions due to the fact that tags in there are more personalized than the other datasets and therefore could not be carried out by WordNet.

On the other hand, a high number of expansions motivated the high percentage of groundings under the *All Expansion* strategy. The *Sibling* and *MFT* strategies in spite of achieving lower rates than *All Expansion* strategy, they reached considerable number of groundings since the context is restricted. Concerning the datasets, Digg and Flickr particularly achieved better grounding rates than Del.ici.ous because in these datasets tags are less personalized as Del.ici.ous is.

Table 3. Semantic expansions and groundings using the twenty ontologies.

| Data | Del.ici.ous | Flickr | Digg |
|---|---|---|---|
| Semantic Expansions | 33% | 32% | 35% |
| All Expansion Strategy | 25% | 19% | 18% |
| Sibling Strategy | 31% | 24% | 19% |
| MFT Strategy | 22% | 18% | 21% |

Table 3 shows the semantic expansions and groundings achieved using the 20 ontologies mentioned in Section 5.1. In general, the whole rates showed in the Table 3 are lower than the rates found in the Table 2. This discrepancy confirms the assumption that the WordNet is broader than the domain ontologies and more tags are expanded and grounded. Different from the values obtained in Table 2, the *All Expansion* strategy did not achieved rates higher than *Sibling* and *MFT* strategies. This is a clear example how expansions and groundings are hard to be undertaken when using ontologies. The three strategies however followed the same behaviour. Concerning the datasets, Del.ici.ous has reached the highest rates comparing to the others datasets. This finding indicates that tags with syntax variations are found in the domain ontologies. In addition, it is reasonable to say that tags in Del.ici.ous are closer to the domains cover by the ontologies than Flickr and Digg.

Table 4. WordNet versus Ontologies.

| Data | WordNet | Ontologies |
|---|---|---|
| Semantic Expansions | 57% | 33.3% |
| All Expansion Strategy | 52.3% | 20.6% |
| Sibling Strategy | 44% | 24.6% |
| MFT Strategy | 38.6% | 20.3% |

Table 4 outlines the means of expansions and groundings from WordNet and Ontologies. The obtained values confirm that WordNet is able to expand tags meaning easier than domain specific ontologies. However it does not necessarily means that WordNet is always advisable or better. The semantic properties in the ontologies are firm representation between two entities while in WordNet is a simply synonymy statement. Moreover, being too generic as WordNet is, it can still keep ambiguity problems unresolved and as a consequence allow the generation of inefficient recommendations. On the other hand, recommendations under ontology grounding are more difficult to be processed and to depend on the quality of the ontologies, semantic expansions and groundings cannot even be performed.

## 5.3. Analyzing Variations of the Recommendations Generated under Semantic Grounding Strategies

In order to calculate the rate of variation, we first generated the recommendations based on cosine similarity in which only lexical comparison is processed regardless semantics. The second step was to generate the recommendations under the predefined strategies and compare them against the previous ones. Table 5 shows the rate of variation to each strategy separately.

Table 5. Variation of the recommendations after semantic grounding.

| Strategy | Del.ici.ous | Flickr |
|---|---|---|
| All Expansion Strategy | 52% | 67% |
| Sibling Strategy | 38% | 41% |
| MFT Strategy | 29% | 32% |
| All Expansion Strategy | 52% | 67% |

Analyzing the values obtained in the Table 5, the rates achieved with *All Expansion* strategy is considerably higher than the other strategies. Notably, more than 50% of the recommendations suffered variations which represent significant changes from last recommendations. Focusing on the datasets, Flickr particularly reached 67% of changes which may be caused due to the existence of ontology devoted exclusively to *photography*. Recommendations from Del.ici.ous and Digg data however kept same behavior of variation considering *All Expansion* strategy.

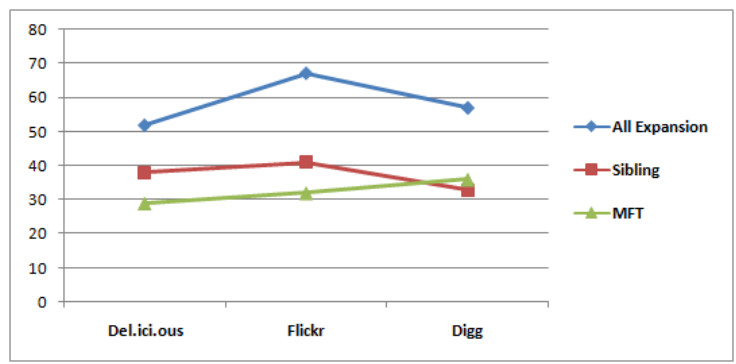

Figure 5: Semantic Expansion Strategies x Dataset

Figure 5 outlines that *Sibling* and *MFT* strategies follow the same behavior achieving lower rates than *All Expansion* strategy. This confirms that the recommendations are restricted by the context when performing the semantic grounding. Comparing both strategies to each other, the recommendations processed under the *MFT* strategy are slightly lower than the *Sibling* strategy. A possible reason for this fact is the difficulty to identify the *MFT* of some users. Looking at each dataset, we observe that Del.ici.ous has reached lowest rates of variation which confirm the hypothesis that tags there are more personalized than the in other datasets. Moreover, the highest rates achieved by Flickr confirm that the existence of ontology in the domain of dataset utilized increases significantly the number of different recommendations.

In spite of that this study demonstrates significant findings, a qualitative experiment should be undertaken in order to confirm the expectation issued here. An experiment in which users expose their satisfaction about the received recommendations should be performed.

## 6. CONCLUSIONS

This paper introduces a study of semantic grounding used for tag similarity calculus. The report shows a comprehensive analysis of semantic grounding given by WordNet dictionary and domain ontologies from different domains. The comparative analysis between these treasures reveals that even though WordNet is more requested, the ontologies provide more concrete semantic expansions during the semantic grounding process. Our study showed that WordNet was able to ground 50% of the tags utilized while domain ontologies only carried out about 30% approximately. Despite Wordnet scores better as it is broader, it does not support several semantic relationships as ontology does. We also realized that personalized tags tend to be easier grounded by domain specific ontologies. Furthermore, the study reveals that even with such number of expansions, the recommendations change significantly. As a future work, it is intended to explore the role of ontology properties in the semantic similarity process and perform a qualitative experiment in which users makes their satisfaction about the received recommendations explicit.

International Journal of Web & Semantic Technology (IJWesT) Vol.2, No.4, October 2011

## ACKNOWLEDGEMENTS

The research leading to these results is part of the project "KiWi - Knowledge in a Wiki" and has received funding from the European Community's Seventh Framework Programme (FP7/2007-2013) under grant agreement No. 211932.

International Journal of Web & Semantic Technology (IJWesT) Vol.2, No.4, October 2011

**Authors**

Frederico Durao graduated with a B.Sc. in Computer Science in 2004. After a period of work experience in the software industry, in 2009 he obtained a Master's Degree in Computer Science at Federal University of Pernambuco. Currently, he is a doctoral student at the department of Computer Science, Aalborg University. He is member of IWIS group (http://iwis.cs.aau.dk) and his research interest includes web personalization, semantic and social web. 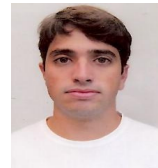

Peter Dolog received the doctoral degree in computer science from the University of Hannover (summa cum laude) in 2006, where he also worked as a researcher from 2002 to 2006. He has been an associate professor of computer science at the Aalborg University since 2008. He studied computer science at the Slovak University of Technology in Bratislava. He heads the IWIS group (http://iwis.cs.aau.dk) and does research in the areas of personalization and recommendation strategies on the Web and Web engineering. He has been a program committee member and reviewer of numerous international conferences and journals and participated in various roles in several collaborative projects at the national, European and international levels. 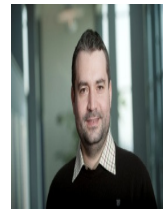